\documentclass[preprint2]{aastex}  
\usepackage{natbib}\usepackage{txfonts}\usepackage{balance}
\usepackage{graphicx}
\usepackage{subfigure}
 
\usepackage{txfonts}

\shorttitle{Dynamic screening correction for solar \emph{p}--\emph{p} reaction rates}
\shortauthors{Mussack \& D\"appen}

\begin{document} 

\title{Dynamic screening correction for solar \emph{p}--\emph{p} reaction rates}

\author{Katie Mussack}
  \affil{Los Alamos National Laboratory, XTD-2, MS T-086, Los Alamos, NM 87545, USA}
  \email{mussack@lanl.gov}

\author{Werner D\"appen}
  \affil{Department of Physics and Astronomy, University of Southern
  California, Los Angeles, CA 90089, USA}

\begin{abstract}

The solar abundance controversy inspires renewed investigations of
the basic physics used to develop solar models. Here we examine the
correction to the proton-proton reaction rate due to dynamic screening
effects. Starting with the dynamic screening energy from the
molecular-dynamics simulations of Mao et al., we compute a
reaction-rate correction for dynamic screening. We find that, contrary
to static screening theory, this dynamic screening does not significantly 
change the reaction rate from that of the bare Coulomb potential. 

\end{abstract}


\section{INTRODUCTION} 

Solar models generated with the \citet{Grevesse_1993} or \citet{Grevesse_1998} abundances
agree quite nicely with helioseismic inferences of the sound speed,
the location of the base of the convection zone, and the helium
abundance in the convection zone. However,
\citet{Asplund_2005,Asplund_2009}, 
\citet{Caffau_2008,Caffau_2009}, and \citet{Ludwig_2009} revised
the solar abundances using three-dimensional hydrodynamic models of the atmosphere with
improved input physics and non-local thermodynamic equilibrium
effects, lowering abundances by up to 1/3. Solar models that use the new lower abundances yield worse
agreement than those that use the older abundances. This has led to years of
heated debates over whether the disagreement represents a solar model
problem or a solar abundance problem. The ongoing disagreement has inspired
a re-examination of all aspects of solar 
models, including opacities, diffusive settling, convective
overshooting, and the possible accretion of low-\emph{Z} material late in the Sun's
evolution or mass loss early in the Sun's evolution \citep[see][for
reviews of mitigation
attempts]{Basu_2008,Guzik_2008,Guzik_2010}. Although these proposed adjustments
have led to some improvement  
in the agreement, they do not satisfactorily resolve the
issue. Further investigations into the basic physics of the solar
interior are required.
   
Nuclear
reactions generate the energy that drives our Sun. Developing an
accurate picture of the conditions that lead to nuclear reactions
is essential in order to fully understand the inner workings of the
Sun. With that in mind, we re-examine screening effects in the solar
core. 
In this work, we focus on \emph{p}--\emph{p}  reactions which
produce most of the nuclear energy generated in the Sun.

\subsection{Nuclear reaction rates}
\label{sect:rates}

In this section, we derive an expression for calculating nuclear reaction rates following the treatment of  \citet{Clayton_1968}. The reaction rate per unit volume between particles of types $\alpha$
and $\beta$ is a product of the number densities of the particles,
$n_{\alpha}$ and $n_{\beta}$, and the average value of the product of
the relative velocity $v$ times the cross section $\sigma$, 
\begin{eqnarray}  
 \label{eq:rate}
  r_{\alpha \beta} & = &
  \frac{n_{\alpha}n_{\beta}}{1+\delta_{\alpha\beta}}\langle\sigma v \rangle_{\alpha
      \beta}  \nonumber  \\
   & = & \frac{n_{\alpha}n_{\beta}}{1+\delta_{\alpha\beta}}
      \int_0^\infty\psi(E)v(E)\sigma(E)dE, 
\end{eqnarray}
where $\psi(E)$ is the relative velocity distribution and $\delta_{\alpha\beta}$ accounts for reactions of like particles. Because we will be dealing with the correction due to dynamic
screening (a ratio between the unscreened and screened reaction
rates), we can ignore the density factor and focus on the reaction rate per pair of particles,
\begin{eqnarray}
 \label{eq:lambda1}
  \lambda &=& \langle\sigma v \rangle_{\alpha \beta}  \nonumber \\
   &=& f(\mu,T)
   \int_0^\infty E\;{\rm{exp}}\left(-\frac{E}{k_BT}\right) \sigma(E)dE,
\end{eqnarray}
where 
\begin{equation}
f(\mu,T) = \sqrt{8 \over {\pi \mu}}\:\left(\frac{1}{k_B T}\right)^{3/2},
\end{equation}
$\mu$ is the reduced mass of the pair, and the Maxwell--Boltzmann distribution is used for 
$\psi(E)$. The cross section $\sigma(E)$ can be
defined as a product of three separate energy-dependent factors
\begin{equation}
 \label{eq:sigma}
  \sigma(E) = \frac{S(E)}{E} \; {\rm{exp}}\left( \frac{-b}{\sqrt{E}}\right),
\end{equation}
where $b= 31.28 Z_{\alpha}Z_{\beta}A^{1/2} \;\rm{keV}^{1/2}$, with $Z_{\alpha}$ and $Z_{\beta}$ being the charges of the interacting ions and $A$
is the reduced atomic weight. The
exponential factor in this expression comes from the barrier
penetration probability, the 
inverse energy dependence comes from the quantum--mechanical
interaction between the two particles, and $S(E)$ contains the
intrinsically nuclear parts of the probability for a nuclear reaction
to occur. With this substitution for $\sigma(E)$ , 
Equation \ref{eq:lambda1} can be re-written as
\begin{equation}
 \label{eq:lambda2}
  \lambda =                        f(\mu,T)
  \int_0^\infty                    
  S(E)\:{\rm{exp}}\left( -{{E} \over {k_B T}} - {{b} \over {\sqrt{E}}} \right) dE.
\end{equation}
In the non-resonant reaction case, $S(E)$ is slowly varying with $E$,
so we can treat it as a constant $S_0$ evaluated at the energy where
${\rm{exp}}(-E/k_BT -b/E^{1/2})$ is maximum. Then the reaction rate per pair of particles (without screening) can be computed as
\begin{equation}
 \label{eq:lambda3}
    \lambda = 
         f(\mu,T)
        S_0 \int_0^\infty 
  {\rm{exp}}\left( -{{E} \over {k_B T}} - {{b} \over {\sqrt{E}}} \right) dE.
\end{equation}

\subsection{Electrostatic screening}
\label{sect:static}

\citet{Salpeter_1954} developed a treatment to include the effect of
static electron screening on nuclear reaction rates. Here we summarize
his method which we will use in Section \ref{sect:method} as the inspiration
for our calculation of the dynamic screening correction. 

We begin by writing the total interaction energy as a combination of the
bare Coulomb potential and a contribution from the plasma: 
\begin{equation}
 \label{eq:int.energy1}
  U_{\rm{total}}(r) = \frac{Z_1Z_2e^2}{r} + U(r).
\end{equation}
Then consider a case in which the classical impact parameter $r_{\rm{c}}$ is
very small compared with the charge cloud radius $R_D$ and the nuclear radius
$r_{\rm{n}}$ is much smaller than $r_{\rm{c}}$. Then the barrier
penetration factor for $r_{\rm{n}}<r<r_{\rm{c}}$ depends only on the expression  
\begin{equation} 
 \label{eq:Edep}
 E - U(r) - \frac{Z_1Z_2e^2}{r} .
\end{equation}
For distances larger than $r_{\rm{c}}$, the barrier penetration factor
hardly depends on the potential. $U(r)$ must be small for distances
greater than $R_D$ and approach a constant value
$U_0$ of the order of magnitude of $Z_1Z_2e^2/R_D$ for small $r$. Then,
\begin{equation} 
 \label{eq:ineq}
 \frac{r_{\rm{c}}}{R_D} \approx \frac{U_0}{E_{\rm{max}}}\ll1,
\end{equation}
where $E_{\rm{max}}$ is the relative kinetic energy for which the integrand in Equation
\ref{eq:lambda1} reaches a sharp maximum. If
this inequality is satisfied, $U(r)$ can be replaced by the potential
at the origin $U_0$.  By examining expression \ref{eq:Edep}, we can see that the screening potential has effectively increased the kinetic energy by a magnitude of $U_0$, so the cross section factors for 
$U_{\rm{total}}=U_{\rm{Coulomb}}+U_0$ are equivalent to the unscreened
factors with energy $E-U_0$. Equation \ref{eq:lambda1} can then be replaced by 
\begin{equation}
 \label{eq:scr.lambda}
  \lambda = 
                      f(\mu,T)
   \int_0^\infty E\;{\rm{exp}}\left(-\frac{E}{k_BT}\right) \sigma(E-U_0)dE.
\end{equation}
With the change of variables $E'=E-U_0$ and the approximation
$(E'+U_0) \approx E'$, the reaction rate
per pair of particles becomes
\begin{equation}
 \label{eq:scr.lambda2}
  \lambda = 
                      f(\mu,T)
   \int_{-U_0}^\infty E'\;{\rm{exp}}\left(-\frac{E'+U_0}{k_BT}\right) \sigma(E')dE'.
\end{equation}
Because the penetration factor for $E'=-U_0$ is so small, the lower limit of the integral can be set to zero without
significantly changing the value of the integral. We then see that
\begin{equation}
 \label{eq:scr.lambda3}
  \lambda_{\rm{screened}}={\rm{exp}}\left(\frac{-U_0}{k_BT}\right)\lambda_{\rm{bare}},
\end{equation}
illustrating that the reaction rate for the statically screened
potential can be approximated by multiplying the rate for the bare Coulomb
potential by ${\rm{exp}}\left(-U_0/k_BT\right)$.

Salpeter then derives $U_0$ by solving the 
Poisson-Boltzmann equation for electrons and ions in a plasma under the 
condition of weak screening ($U_{\rm{total}}(r)\ll k_BT$). He 
arrives at an expression for the screening energy that is equivalent to 
that of the Debye--H\"uckel theory of dilute solutions of electrolytes 
\citep{Debye_1923}:
\begin{equation}
 \label{eq:Debye}
  U_0 = -\frac{Z_1Z_2e^2}{R_D}. 
\end{equation}
The Debye length, $R_D$, is the characteristic screening length of
a plasma at temperature $T$ with number density $n$  which is defined by
\begin{equation}
 \label{eq:length_general}
  R_D^2 = \frac{\epsilon_0 k_B T}{e^2(n_e+n_iZ_i^2)}. 
\end{equation}
For a neutral proton--electron plasma, the electron number density $n_e$ and ion number density $n_i$ are both $n/2$, so the Debye length is just 
\begin{equation}
 \label{eq:length}
  R_D^2 = \frac{ \epsilon_0 k_B T}{ne^2}. 
\end{equation}

\subsection{Dynamic screening}
\label{sect:dynamic}

Although Salpeter's expression accurately describes the effect of
static screening, the issue of dynamic screening in the hot, dense plasma
of the solar interior remains an open question. Dynamic screening
occurs when the screened interaction energy of a pair of ions depends
on the relative velocity of the pair. Most of the ions in the solar plasma are
much slower than the electrons and the fastest ions. The thermal ions are therefore not able to
rearrange themselves as quickly around individual faster moving
ions. Since nuclear reactions require energies several times the
average thermal energy, the ions that are able to engage in nuclear
reactions in the Sun are the faster moving ions, which are not
accompanied by a full static screening cloud. 

Salpeter's derivation uses the mean-field approach in which 
the many-body interactions are reduced to an average interaction that
simplifies calculations. This technique is quite useful for
calculations describing the average behavior of the plasma. However,
dynamic effects for the fast-moving, interacting ions in hot, dense
plasma lead to a
screened potential that deviates from the average value. Therefore,
the mean-field approximation is not appropriate for computing stellar nuclear
reaction rates. Instead, we use the molecular-dynamics method of 
\citet{Shaviv_1996} to model the motion of
protons and electrons in a plasma under solar conditions in order to
investigate dynamic screening in \emph{p}--\emph{p} reactions. The advantage of
the molecular-dynamics method is that it does not assume a mean
field. Nor does it assume a long-time average potential for the
scattering of any two charges, which is necessary in the statistical
way to solve Poisson's equation to obtain the mean potential in a plasma.  

\begin{figure}[ht]
 \begin{center}
  \includegraphics[width=82mm]{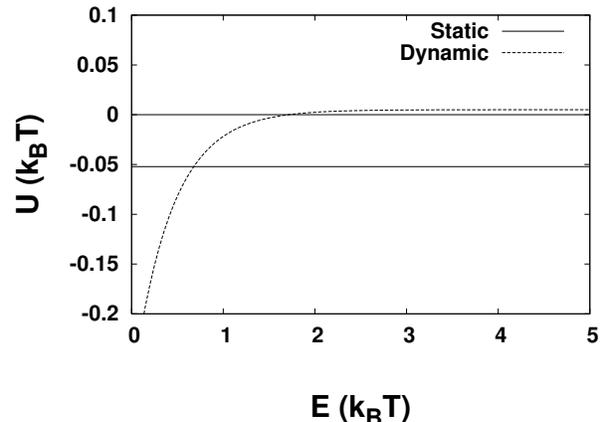}
  \vskip -0.4cm \caption{Dynamic screening energy at the turning point for pairs of protons with a given relative kinetic energy. For comparison, the
static screening energy evaluated at the average turning point of proton
pairs with each energy is also shown. Data from \citet{Mao_2009}.} 
  \label{fig:Usim} \vskip .3cm
 \end{center}
\end{figure}

In previous work, \citet{Mao_2009} present
simulation results for the velocity-dependent screening energy of
\emph{p}--\emph{p} reactions in a plasma with the temperature and density of the
solar core ($T=1.6 \times 10^7$  \rm{K}, $\rho = 1.6 \times 10^5$
$\rm{kg \; m^{-3}}$). They demonstrate that the static screening result does 
not accurately represent this plasma, and they compute a screening energy
that depends on the relative kinetic energy of a pair of interacting
ions (see Figure \ref{fig:Usim}). In this paper, we use their simulation results to compute a
correction to the solar \emph{p}--\emph{p} reaction rate due to the dynamic screening they observe.

\section{METHODS}
\label{sect:method}

We begin with the calculations of \citet{Mao_2009} for a plasma of
protons and electrons with the temperature and density of the solar
core. Their Figure 5 shows the relationship between the total interaction
energy at the turning point and the relative kinetic energy of a pair
of interacting protons. As in Salpeter's static screening derivation, we can split the total interaction
energy into the Coulomb and screening cloud contributions:
\begin{equation}
 \label{eq:int.energy2}
  U_{\rm{total}}(r) = \frac{Z_1Z_2e^2}{r} + U(r,v).
\end{equation}
The screening energy now includes a velocity dependence, so $U_0$ can no
longer be factored out of the energy integral as was done to obtain
Equation \ref{eq:scr.lambda3} for the static screening case. 
 
Following
Salpeter's calculation for static screening, we focus on the
contribution to the interaction energy from the screening cloud at
small $r$. This value of $U_0$ is obtained from the relationship between the
screening energy of a pair of protons at their turning point and their
relative kinetic energy  which is shown in Figure \ref{fig:Usim}.

The dynamic screening energy curve is described by the equation
\begin{equation}
 \label{eq:Uofv}
 \frac{U_0(E)}{k_BT} = 0.005 - 0.281\;{\rm{exp}}\left(-2.35 \frac{E}{k_BT}\right) ,
\end{equation}
which comes from the best-fit curve for the \citet{Mao_2009}
$E_{\rm{screen}}(r_{\rm{c}},E)$. (Note the difference in sign from
\citet{Mao_2009}, where the screening energy was defined as a negative
contribution to the total energy,
$E_{\rm{total}}(r)= e^2/r-E_{\rm{screen}}(r,E)$. In this paper, we use
the \citet{Salpeter_1954} convention, as shown in Equations
\ref{eq:int.energy1} and \ref{eq:int.energy2}.) 
  
Now we return to Equation \ref{eq:scr.lambda} for the screened
reaction rate per pair of particles. The assumptions and approximations used in the derivation of Equation \ref{eq:scr.lambda} for the static case will be examined and justified for the dynamic case in Section \ref{subsect:assumptions}.  Replacing $U_0$ from the
statically screened case with $U_0(E)$ for the dynamically screened
case and using definition \ref{eq:sigma} for the cross section, we have
\begin{eqnarray}
 \label{eq:dyn.lambda} 
  \lambda = 
                      f(\mu,T)
   \int_0^\infty  \frac{E}{E-U_0(E)}\;  S(E-U_0(E))  \nonumber \\
  \quad  \times \;{\rm{exp}}\left[-\frac{E}{k_BT}-\frac{b}{\sqrt{(E-U_0(E))}} \right]  dE.
\end{eqnarray}
 Because $S(E)$ is a slowly varying function of energy, we make
 the approximation 
\begin{equation}
 \label{eq:Sapprox}
  S(E-U_0(E)) \approx S(E)
\end{equation}
 and replace this function with the constant $S_0$, as was done in the original
 reaction-rate calculation in Equation \ref{eq:lambda3}.  
 
We can now evaluate the reaction rate per pair of particles using
Equation \ref{eq:lambda3} for the unscreened case and
\begin{eqnarray}
 \label{eq:dyn.lambda2}
  \lambda  =                         f(\mu,T)
   S_0  \nonumber   \\
   \times \int_0^\infty \frac{E}{E-U_0(E)}  
   \; {\rm{exp}}\left[-\frac{E}{k_BT}-\frac{b}{\sqrt{(E-U_0(E))}}
    \right]dE  
\end{eqnarray}
for the statically and dynamically screened cases. Equation \ref{eq:Debye} gives the static
screening $U_0$ and Equation \ref{eq:Uofv} gives the dynamic screening
$U_0(E)$. Because all three cases contain the factor $
  f(\mu,T)
  S_0
  $, we only need to
compute the integrals in order to compare ratios of the two screened cases to
the bare Coulomb potential case to obtain the correction factors for
the \emph{p}--\emph{p} reaction rate.

\section{RESULTS}

\begin{table*}
 \begin{center}
  \begin{tabular}{|l|c|c|}
    \hline
     Case & Screening energy U & Reaction-rate correction   \\ \hline \hline
     Unscreened  & 0  & 1        \\ \hline
     Statically screened & $U_0 = -\frac{Z_1Z_2e^2}{R_D}$  & 1.042   \\ \hline
     Dynamically screened  & $U_0(E)  = k_BT\left( 0.005 - 0.281\;{\rm{exp}}\left(-2.35 \frac{E}{k_BT}\right)\right) $ & 0.996 \\ \hline
   \end{tabular}
  \end{center}
 \caption{Screening energies and the ratio of screened to unscreened nuclear reaction rates for solar p-p reactions.}
 \label{tab:integrals}
\end{table*}

Table \ref{tab:integrals} shows the results of the screening
corrections for solar \emph{p}--\emph{p} reaction rates computed from the integrals in Equations
\ref{eq:lambda3} and \ref{eq:dyn.lambda2}.  The statically
screened correction shows a fairly large enhancement in the nuclear reaction
rate. Conversely, the dynamically screened reaction rate is almost the
same as the unscreened rate.

\subsection{Integrands}

How does the dynamic screening energy seen in Figure \ref{fig:Usim} result in a reaction rate that is so close to the unscreened reaction rate? To answer that question, we compare the components of the the integrands from Equations \ref{eq:lambda3} and \ref{eq:dyn.lambda2}. Both integrands can be written in the general form
\begin{equation}
 \label{eq:FE}
	F(E) = H(E) J(E),
\end{equation}
where 
\begin{equation}
 \label{eq:HE}
	H(E) = \frac{E}{G(E)},
\end{equation}
\begin{equation}
 \label{eq:JE}
	J(E) = {\rm{exp}}\left[-\frac{E}{k_BT}-\frac{b}{\sqrt{G(E)}},
    \right]
\end{equation}
and
\begin{equation}
 \label{eq:GE}
	G(E) = E-U.
\end{equation}
The three cases only differ in the screening energy $U$ which is shown for each case in Table \ref{tab:integrals}. 

\begin{figure*}[ht]
 \begin{center}
   \subfigure[$G(E)$] { \label{fig:GE}  \includegraphics[width=80mm]{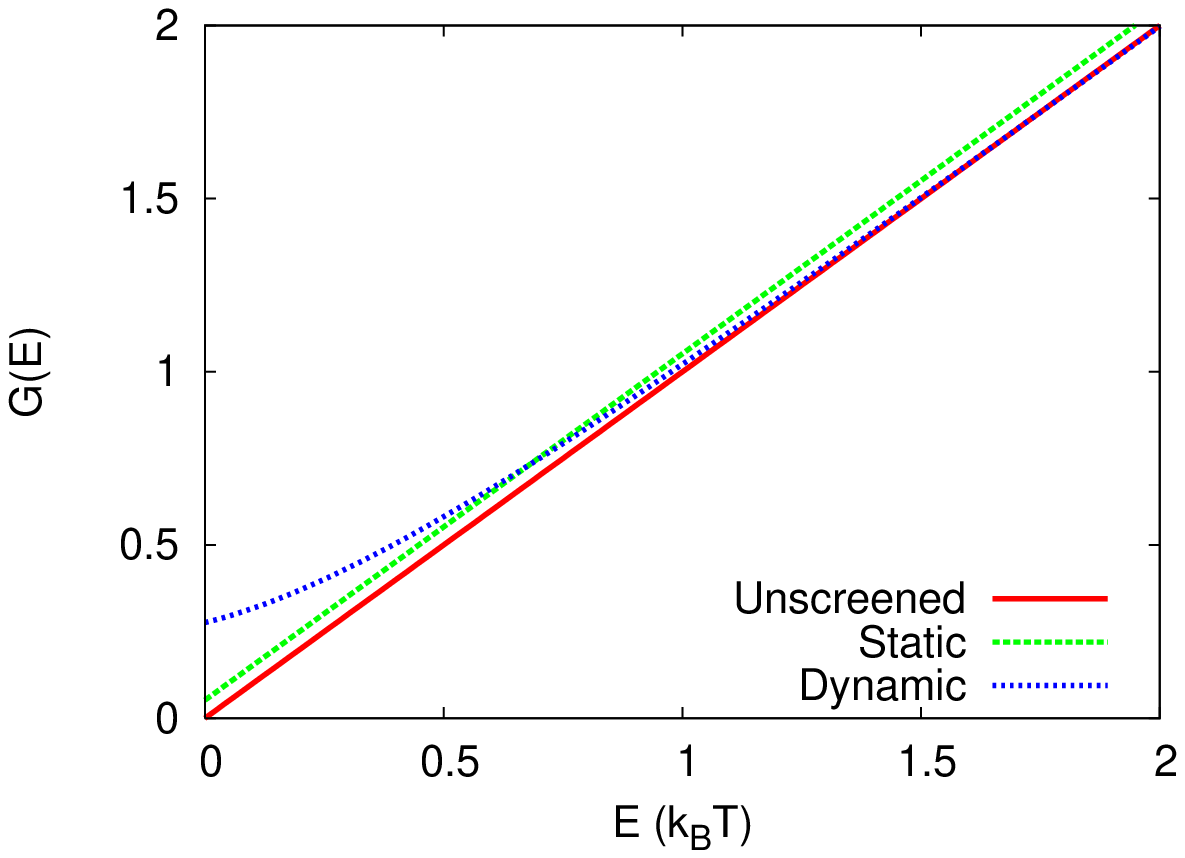}}
   \subfigure[$H(E)$] {\label{fig:HE}  \includegraphics[width=80mm]{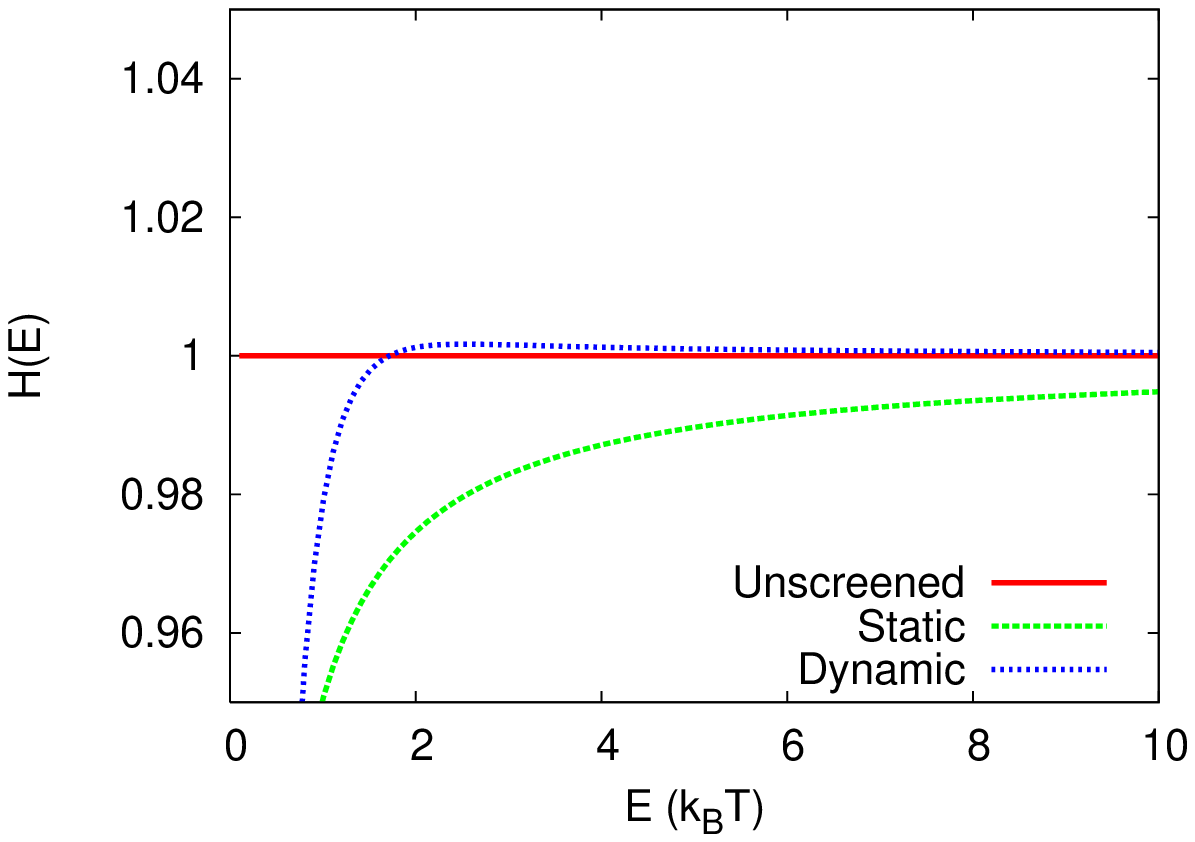}}
   \subfigure[$J(E)$] {\label{fig:JE}  \includegraphics[width=80mm]{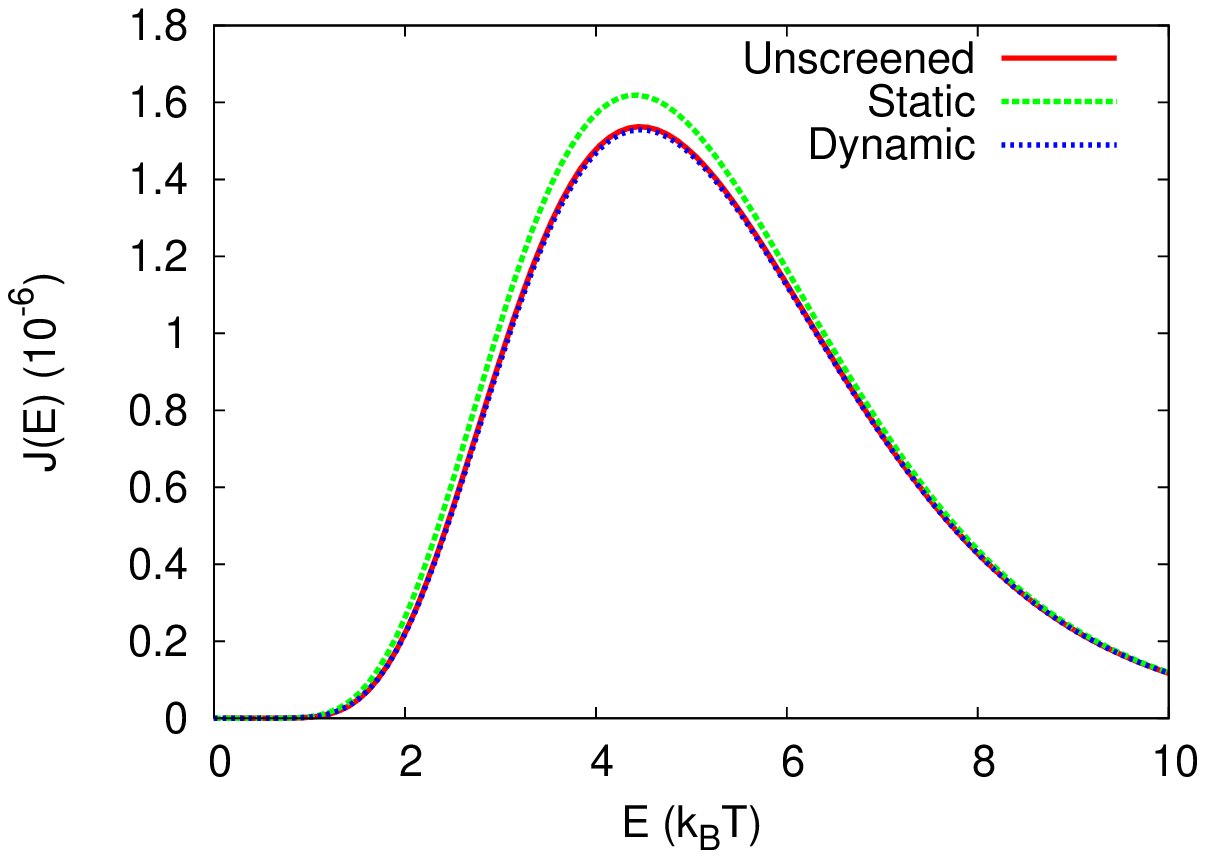}}
   \subfigure[$F(E)$ ] {\label{fig:FE}  \includegraphics[width=80mm]{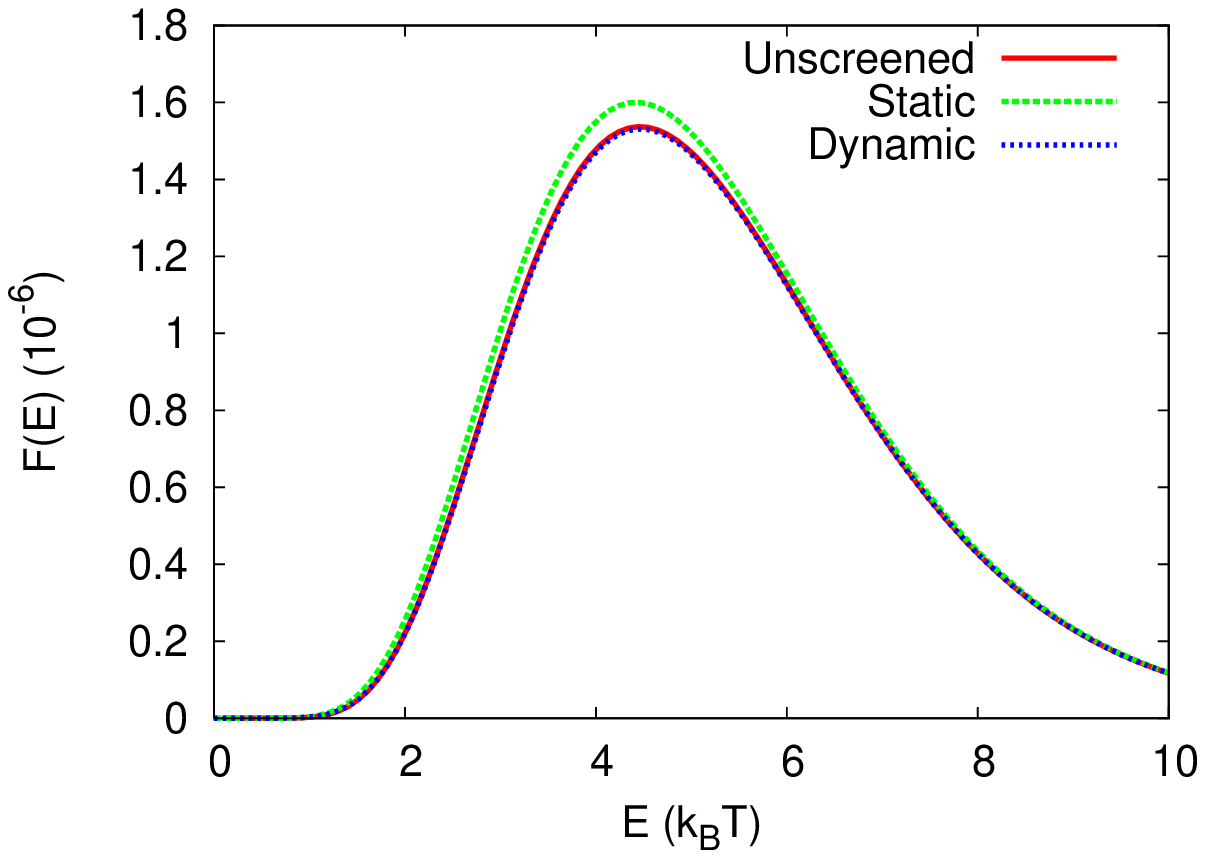}}
  \end{center}
\caption{The functions $G(E)$, $H(E)$, $J(E)$, and $F(E)$ from equations  \ref{eq:GE},  \ref{eq:HE},  \ref{eq:JE},  and \ref{eq:FE} for the unscreened, statically screened, and dynamically screened cases.}
\label{fig:integrand}
\end{figure*}

In Figure \ref{fig:GE}, we see that the dynamic $G(E)$ approaches the unscreened $G(E)$ for high energies. Figure \ref{fig:HE} shows that this leads to the dynamic $H(E)$ approaching the unscreened $H(E)$ for energies above $\sim$2 $k_BT$, while the static $H(E)$ is very different from both the dynamic and bare $H(E)$. Below energies of $\sim$2 $k_BT$ the dynamic $H(E)$ drops rapidly away from the unscreened value of 1. However, the factor $H(E)$ is multiplied by $J(E)$ in the integrand of the reaction-rate equations. As seen in Figure \ref{fig:JE}, $J(E)$ is very small to zero below energies of $\sim$2 $k_BT$, damping out the region of $H(E)$ in which the unscreened and dynamic results diverge. This leads to integrands for the dynamic and unscreened cases that are nearly identical, as seen when $H(E)$ and $J(E)$ are multiplied together to give $F(E)$ in Figure \ref{fig:FE}. The Gamow-peak-like factor $J(E)$ acts as a weighting function to devalue the $H(E)$ contribution of the slow pairs of ions that rarely participate in nuclear reactions. The faster pairs that cause less polarization of the surrounding plasma and therefore see less screening provide the main contribution to the reaction-rate integral.

\subsection{Evaluating assumptions}
 \label{subsect:assumptions}
 
Now that we have defined the integrand $F(E)$ for the dynamic case, we can return to the issue of assumptions and approximations that were justified in the static screening rate derivation and adopted in the dynamic screening rate derivation. Here we assess the validity of these assumptions and approximations in the case of dynamic screening. 

The first assumption is that $r_{\rm{c}}$ is very small compared with the charge cloud radius $R_D$ and that the nuclear radius $r_{\rm{n}}$ is much smaller than $r_{\rm{c}}$, leading to expression \ref{eq:Edep}. Although the dynamic case does not have a traditional static screening cloud, the inequality $r_{\rm{n}}\ll r_{\rm{c}}\ll R_{\rm{D}}$ is still satisfied. This can be seen in the definition of $R_{\rm{D}}$ as the distance beyond which an appreciable fraction of the nuclear charge is screened by the polarization charge cloud, a distance that is large for the dynamic case.

Next we examine the inequality in Equation \ref{eq:ineq}, $U_0/E_{\rm{max}}\ll 1$. In the dynamic case, $E_{\rm{max}}$ is the kinetic energy for which the integrand of Equation \ref{eq:dyn.lambda2} is maximum. We see from Figure \ref{fig:FE} and $U_0$ in Figure \ref{fig:Usim} that the peak of the integrand $F(E)$ occurs at energies for which $U_0/E_{\rm{max}}\ll 1$.

Finally, we address the approximation that $U(r)$ can be replaced by the potential at the origin $U_0$. While this is easy to see for the static case, in the case of dynamic screening we do not have the screening energy in the form $U(r,E)$ to examine this claim. Instead, we define $U_0(E)$ to be the screening energy computed at the turning point of the approaching protons, since this is the relevant $U(r,E)$ for nuclear reaction-rate calculations.

\section{DISCUSSION OF ARGUMENTS AGAINST DYNAMIC SCREENING}

In light of the contentious debate over the validity of dynamic screening, we devote this section to a discussion of  arguments that have been made against dynamic screening.

\subsection{Incorrect derivations}

The argument for dynamic screening has been damaged by several derivations of alternate screening formulae \citep[see, for example,][]{Carraro_1988, Opher_2000, Shaviv_1996, Tsytovich_2000} that were subsequently shown to be incorrect. Here we discuss two examples, \citet{Carraro_1988} and \citet{Shaviv_1996}.  \citet{Carraro_1988} derived a modified screening potential for fusing ions when the Gamow velocity is greater than the thermal velocity. \citet{Brown_1997} showed that including processes of excitation or de-excitation of the plasma in an interaction with one of the fusing ions exactly cancels the dynamic modifications proposed by \citet{Carraro_1988}. \citet{Shaviv_1996} then introduced a factor of 3/2 on the screening energy. They arrived at this result by including the interaction of the the screening cloud from each fusing ion in the total interaction potential. \citet{Bruggen_1997} showed that \citet{Shaviv_1996} misinterpreted the thermodynamics and used an incorrect potential in the Schr\"odinger equation for the system. \citet{Bahcall_2002} summarize the problems with several different alternative screening formulae. However, finding flaws in these (and other) derivations of analytical expressions for screening deviations does not rule out the effect of dynamic screening. This argument only highlights the difficulty in developing a general analytical formalism to describe dynamic screening.

\subsection{Factorability of the distribution function is wrong}

For the Gibbs distribution, probabilities for momenta and coordinates are independent and cannot influence each other. This leads to the argument that velocities of fusing particles cannot have an effect on screening because the distribution function is not factorable.  However, it is clear when examining individual ions that their relative velocity affects how close the ions can be to each other. This argument extends to ions in a plasma, where the configuration of the screening cloud of approaching ions depends on the relative velocity of those ions.  Over the whole system, this velocity-dependent effect averages out to the Gibbs distribution, but each screening cloud is not identical to the average configuration of a screening cloud in that system.

Solar nuclear reactions select a biased sample of the ions in the system. These nuclear reactions involve mainly the fastest ions, not a random sample of all ions in the system. Therefore the velocity distribution must be multiplied by the velocity-dependent screening energy before integration instead of beginning with an average value of the distribution.

\subsection{Higher-order terms}

Do dynamic screening results imply that higher-ordered terms are required? The screening energy can be expressed as a power series expansion in the plasma coupling parameter $\Lambda$:
\begin{equation}
 \label{eq:expansion}
   \frac{U(r)}{k_BT} = -\frac{\Lambda}{x}  \left( 1 - \exp \left(-x \right) \right) - \Lambda^2 f(x,\Lambda) \;, 
\end{equation}
where 
\begin{equation}
  \Lambda = \frac{Z_1Z_2e^2}{R_Dk_BT} \;,
\end{equation}
$x=r/R_D$, and $f(x,\Lambda)$ is given by \citet{DeWitt_1965}. The first term reduces to the Debye--H\"uckel weak screening result shown in Equation \ref{eq:Debye}. 

For the temperature and density of our simulations ($T=1.6 \times 10^7$  \rm{K}, $\rho = 1.6 \times 10^5$ $\rm{kg \; m^{-3}}$), $\Lambda = 0.05$, so higher-order terms are small. Therefore, if dynamic screening could be described by higher-order terms in the expansion, the effect should be much smaller than the first term. However, the dynamic correction is not just a higher-order term in the expansion. Dynamic effects come from a different approach to determining screening effects. Instead of deriving an expression for screening based on average properties of the system, we examine the formation of the screening clouds themselves and do not average out the velocity-dependent nature of the clouds.

\subsection{Observational confirmation}

What observational evidence can confirm any screening effect in the nuclear reactions in the Sun? Many early discussions of dynamic screening were motivated by the neutrino problem which has since been resolved with neutrino oscillation theory.  In addition, including dynamic screening corrections in models with the \citet{Grevesse_1993} or \citet{Grevesse_1998} abundances worsened agreement with helioseismic constraints \citep[see, for example, ][]{Weiss_2001}. However, the solar abundance problem provides renewed motivation for exploring dynamic screening in solar nuclear reactions. 

Before 2005, solar models with the latest input physics reproduced the sound-speed profile determined from helioseismic inversions to within $0.4\%$ and also provided good agreement with the seismically inferred convection zone depth and convection zone helium abundance. Then \citet{Asplund_2005,Asplund_2009}, \citet{Caffau_2008,Caffau_2009} and \citet{Ludwig_2009} began using three-dimensional hydrodynamic models of the solar atmosphere with improved input physics and non-local thermodynamic equilibrium effects to determine solar atmospheric abundances. These revised calculations lowered element abundances by up to 1/3. When the lower abundances are incorporated in solar models, the sound-speed profiles, convection zone depths, and convection zone helium abundances give worse agreement with helioseismic constraints than models with the old, higher abundances. Many attempts have been made to improve agreement by adjusting the physics or evolutionary assumptions in solar models  \citep[see][]{Basu_2008, Guzik_2008, Guzik_2010}. Although these adjustments have shown some improvement, no model using the new lower abundances agrees as well with the helioseismic constraints as the models using the older abundances.

\begin{figure}[ht]
 \begin{center}
  \includegraphics[width=82mm]{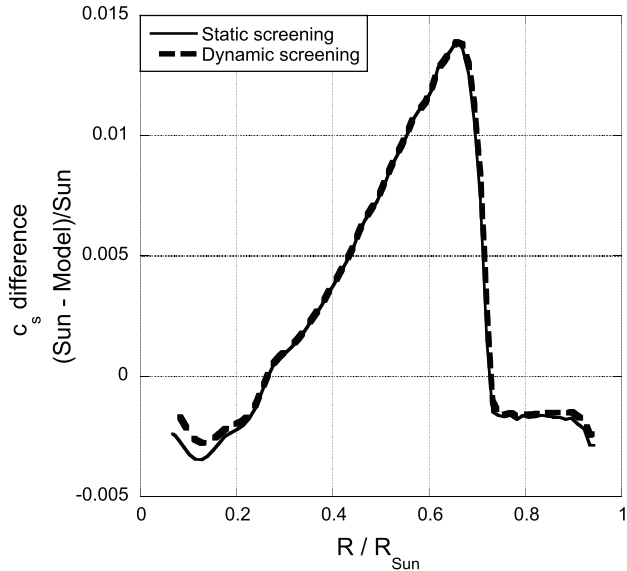}
  \vskip -0.4cm \caption{Difference between inferred and calculated sound speeds for models with the \citet{Asplund_2005} abundances with and without the dynamic screening correction for p-p reaction rates} 
  \label{fig:sound_pp0} \vskip .3cm
 \end{center}
\end{figure}

In a forthcoming paper, \citet{Mussack} incorporate the dynamic screening correction shown here for solar \emph{p}--\emph{p} reaction rates into solar models with the new lower abundances. They show that including this correction in solar models improves the sound speed discrepancy in the solar core, as shown in Figure \ref{fig:sound_pp0}. This improvement does not fully reconcile the new abundances with helioseismic constraints, but it is a step in the right direction. Perhaps in combination with other changes, dynamic screening corrections could contribute to a solution to the solar abundance problem.

\section{SUMMARY}

We have shown that dynamic screening in solar \emph{p}--\emph{p} reactions does not
reproduce the enhancement of reaction rates that is predicted by
Salpeter's static screening approximation. In fact, the dynamic screening seen by 
\citet{Mao_2009} shows essentially no correction to the unscreened reaction rate.  

Although \emph{p}--\emph{p}
reactions in the core are the main source of nuclear energy generated 
in the Sun, this reaction-rate correction is only the beginning of understanding how including
dynamic effects will alter a full solar model. The reaction-rate
correction must be generalized to treat other temperatures, densities,
compositions, and reactions. In addition, the effect of dynamic
screening on the equation of state must be examined. Until we can meet both of
these challenges, dynamic screening cannot be incorporated completely and
consistently in solar and stellar models.

\begin{acknowledgments}

We thank Hugh DeWitt for useful discussions and Dan Mao for her
simulation results. This work was supported in part by grant
AST-0708568 of the National Science Foundation.  

\end{acknowledgments}

\end{document}